\documentstyle[a4,12pt]{article}

\begin{document}
\title{Information theory in the study of anisotropic radiation}
\author{Raquel Dom\'\i nguez-Cascante\thanks{%
Electronic address: raquel@ulises.uab.es}\\
Departament de F\'{\i}sica\\
Universitat Aut\`onoma de Barcelona\\
08193 Bellaterra, Spain}
\date{}
\maketitle
\newpage
\begin{abstract}
Information theory is used to perform a thermodynamic study of non
equilibrium anisotropic radiation. We limit our analysis to a second-order
truncation of the moments, obtaining a distribution function which leads to
a natural closure of the hierarchy of radiative transfer equations in the
so-called variable Eddington factor scheme. Some Eddington factors appearing
in the literature can be recovered as particular cases of our two-parameter
Eddington factor. We focus our attention in the study of the thermodynamic
properties of such systems and relate it to recent nonequilibrium
thermodynamic theories. Finally we comment the possibility of introducing a
nonequilibrium chemical potential for photons.
\end{abstract}

\newpage

\section{Introduction}

The study of radiation hydrodynamics \cite{Pomraning} has proven to be of
great interest in astrophysics, cosmology and plasma physics. The radiative
transfer equation for the specific radiation intensity $I(\vec{r},t,\nu ,%
\vec{\Omega})=h\nu c\ n(\vec{r},t,\nu ,\vec{\Omega})$, where $n(\vec{r}%
,t,\nu ,\vec{\Omega})$ is the occupation number of photons with frequency $%
\nu $ moving in direction $\vec{\Omega}$, is in many practical situations
too involved to be solved analytically. What is usually done is to consider
the equations for the moments of $I(\vec{r},t,\nu ,\vec{\Omega})$ up to a
given order $m$ \cite{Lever,Romano}. However, due to the dependence of the
equation for the moment $m$ on the moment $m+1,$ one needs to introduce a
closure relation. If only the energy density $e$ ($m=0$) and the energy flux 
$\vec{J}_E$ ($m=1$) are considered, one must introduce a closure relation
for the pressure tensor ${\bf P}_E$ ($m=2$). Hence, in this approximation
the relevant physical quantities are the angular moments of the intensity
(note that in the following we consider as variables the moment of the
photons $\vec{p}=\,\vec{\Omega}\,h\nu /c=p\vec{c}/c$ instead of the
frequency $\nu $ and the solid angle $\vec{\Omega}$)
\begin{equation}  \label{e}
e(\vec r,t)=\frac 1{h^3}\int pc\,n(\vec r,\vec p,t)d^3\vec p,
\end{equation}
\begin{equation}  \label{Je}
\vec J_E(\vec r,t)=\frac 1{h^3}\int pc\vec c\,n(\vec r,\vec p,t)d^3\vec p,
\end{equation}
\begin{equation}
{\bf P}_E(\vec{r},t)=\frac 1{h^3}\int \vec{p}\vec{c}\,n(\vec{r},\vec{p},t)d^3%
\vec{p},  \label{Pe}
\end{equation}
namely the energy density, the energy flux and the pressure tensor. It is
also convenient to define the following normalized quantities 
\begin{equation}
\vec{f}_E=\frac{\vec{J}_E}{ec},\;\;\;{\bf T}_E=\frac{{\bf P}_E}e.
\label{reduced}
\end{equation}
The closure relation is performed by the introduction of the so-called
Eddington factor $\chi (f_E)$, defined as the eigenvalue of the pressure
tensor corresponding to the eigenvector $\,\vec{n}$ (unitary vector in the
direction of the energy flux), i.e. 
\begin{equation}
{\bf T}_E\,\vec{n}=\chi \,\vec{n}.
\end{equation}
This definition leads to the relation 
\begin{equation}
{\bf T}_E=\frac{1-\chi }2{\bf I+}\frac{3\chi -1}2\vec{n}\vec{n},
\label{Edfact}
\end{equation}
where ${\bf I}$ is the identity matrix. In the limit of isotropic radiation
(Eddington limit) $\chi (0)=1/3$, while in the free streaming case $\chi
(1)=1$.

In addition, one can define in an analogous way the angular moments of the
occupation number $n(\vec r,\vec p,t)$, namely the total photon number
density, the photon flow and the flux of the particle flow 
\begin{equation}  \label{densitat}
\rho (\vec r,t)=\frac 1{h^3}\int_0^\infty \,n(\vec r,\vec p,t)d^3\vec p,
\end{equation}
\begin{equation}  \label{fluxN}
\vec J_N(\vec r,t)=\frac 1{h^3}\int_0^\infty \vec c\,n(\vec r,\vec p,t)d^3%
\vec p,
\end{equation}
\begin{equation}
{\bf P}_N(\vec r,t)=\frac 1{h^3}\int_0^\infty \vec c\vec c\,n(\vec r,\vec p%
,t)d^3\vec p,
\end{equation}
and the corresponding normalized quantities 
\begin{equation}  \label{reducedn}
\vec f_N=\frac{\vec J_N}{\rho c},\;\;\;{\bf T}_N=\frac{{\bf P}_N}{\rho c^2}.
\end{equation}

Whenever the angular and the frequency dependence of the radiation intensity
(or the occupation number) factorize, the two sets of moments are not
independent, but verify the relations
\begin{equation}
\vec{f}_E=\vec{f}_N,\;\;\;{\bf T}_E={\bf T}_N.
\end{equation}
This fact is implicitly assumed in most papers on the subject \cite{Fu}. In
many instances, however, the frequency is not integrated \cite{Lever,Minerbo}
and, thus, in this situation the two sets of moments also coincide. However,
all the previous quantities would be frequency-dependent, except, again, in
the case that radiation intensity factorizes into a frequency-dependent part
and an angular-dependent part.

There is a great amount of different Eddington factors in the literature
introduced following physically different approaches (see \cite{Lever} for a
review). Among the different approaches to obtain variable Eddington
factors, some authors have used a maximum entropy principle, both from a
macroscopic \cite{Anile,Kremer} and from a microscopic point of view \cite
{Fu,Minerbo,Inf}. From a macroscopic viewpoint, balance equations for the
energy density and the energy flux have been considered, and an entropy
principle has been introduced to exploit these constraints, leading to the
so-called Lorentz's Eddington factor that Levermore \cite{Lever} had
previously obtained. In \cite{Inf} the same result was obtained from the
point of view of information theory with fixed energy flux. However, as
pointed out in \cite{MiniCuca}, this anisotropic Eddington factor
corresponds to a Lorentz transformation of the equilibrium one so that the
anisotropy is only due to the fact that the system is being observed from a
moving reference frame. Thus, this Eddington factor does not describe a real
out of equilibrium situation.

On the other hand, both Minerbo \cite{Minerbo} and Fu \cite{Fu} have
obtained different Eddington factors, considering as a constraint the photon
flux instead of the energy flux. In this case, a reference frame where the
anisotropy disappears can not exist.

In this paper, we apply information theory to generalize the two situations
mentioned above, by considering simultaneously the energy flux and the
photon flow as independent variables. Hence, the previous results can be
recovered as limiting cases, while some new situations can be analyzed. In
addition, information theory allows a complete thermodynamic study of
radiation out of equilibrium. The closure scheme previously described, in
which the pressure tensor is written as a function of the energy flux,
departs from the hypothesis of local equilibrium, which implies that the
distribution in momentum space is locally (i.e., for each position) the same
as an equilibrium distribution. However, radiation usually has a
distribution markedly different from a black-body distribution, and this
hypothesis must, therefore, be abandoned in radiative transfer problems \cite
{Essex}. The structure of classical nonequilibrium thermodynamics (for
example, bilinear forms for the entropy production rate) also presents a
lack of consistency for radiation \cite{Essex}. When the radiation field is
strongly anisotropic, the mean free path of the photons is large and the set
of macroscopic quantities describing the local state of the photon gas
arises from interactions occurring over large regions, whereas when the
photon gas is in equilibrium the interactions which thermalize photons and
matter take place in a specific volume. Thus, to describe the radiation gas,
the appropriate set of quantities must contain information about its angular
distribution \cite{Fu}.

This can be done in the closure scheme described above by including the
fluxes among the set of thermodynamic variables with the help of information
theory. This formalism can be used as an heuristic method to find a
distribution function consistent with the information available about the
system. We will be able to analyze also the influence of the dissipative
fluxes in the nonequilibrium equations of state. This procedure to study
nonequilibrium equations of state has already been used in \cite
{Inf,Nettleton,Corbet} in the case of an ideal gas and in \cite{Lebon} to
study heat conduction in a boson gas. Indeed, from the point of view of
nonequilibrium thermodynamics, the meaning of the fundamental thermodynamic
quantities in nonequilibrium states is a basic challenge, so that it
deserves attention from all possible points of view.

The plan of the paper is as follows. In Section 2, we apply the information
theoretical formalism to the study of radiation (within the low-occupation
number approximation) under an energy and a photon flux, considered as
independent variables. Hence, we can obtain a two-parameter Eddington factor
depending on both fluxes and generalized equations of state. We can also
recover Lorentz's and Minerbo's Eddington factors in the proper limits. In
Section 3 we analyze the possibility of introducing a nonequilibrium chemical
potential for photons and its physical consequences. The last section is
devoted to review the main conclusions of the paper.

\section{Anisotropic radiation under energy and particle fluxes}

Information theory was introduced in 1957 by Jaynes in statistical mechanics 
\cite{Jaynes,Llibre} in order to provide a probabilistic basis to
equilibrium thermodynamics. However, although the foundations of the use of
the informational entropy functional for nonequilibrium situations are far
from being trivial, it has also been applied to nonequilibrium situations
(see \cite{Luzzi} and references therein). The method asserts that the
steady state of a system, defined by the values of a set of macroscopic
constraints, is the most (microscopically) disordered state compatible with
these constraints, while disorder is measured by means of Shannon's entropy,
which is defined as follows: let ${\cal N}$ be the number of microstates
compatible with the macroscopic constraints acting on the system and $p_i$
the probability of a given microstate $i$. The informational entropy is
given by: 
\begin{equation}
S=-k_B\sum_{i=1}^{{\cal N}}p_i\ln p_i,  \label{Shannon}
\end{equation}
where $k_B$ is the Boltzmann's constant and providing that the normalization
condition $\sum_ip_i=1$ is fulfilled. By maximizing (\ref{Shannon}) subject
to the constraints (mean values of extensive quantities controlled in a
given experiment) the probability of each microstate is obtained. This
method realizes a probability assignment which is as unbiased as possible
and avoids any unwarranted assumption beyond the information contained in
the constraints. The probabilities $p_i$ corresponding to equilibrium
ensembles can be easily derived from this postulate of maximum entropy as
shown in \cite{TermoCallen}.

In nonequilibrium situations, one considers that the generalized entropy
functional (\ref{Shannon}) depends both on the equilibrium and
nonequilibrium constraints acting on the system and the corresponding
probabilities $p_i$ are obtained by the maximization of this entropy. These
nonequilibrium constraints may be, for instance, the heat flux or the
viscous traceless pressure tensor.

If quantum systems are considered, the statistical entropy can be calculated
in terms of the occupation number $n_i$ according to \cite{Lands2} 
\begin{equation}  \label{Shannon2}
S=k_B\sum_{i=1}^{{\cal N}}\left[ n_i\ln (\frac{g_i}{n_i}-a)-\frac{g_i}a\ln
(1-a\frac{n_i}{g_i})\right] ,
\end{equation}
where $g_i$ is the degeneracy, and $a=0,+1,-1$ for classical particles,
fermions and bosons, respectively.

Our purpose in this paper is to generalize some previous works \cite
{Fu,Minerbo,Anile,Kremer} by means of information theory and derive a more
general form for the Eddington factor. We will study the case of a radiation
gas in the low occupation number approximation (in order to obtain
analytical expressions) submitted both to an energy flux an a particle flux.
The reason is that, in order to obtain a non purely advective energy flux,
as in the Lorentz case and to generalize the study performed by Minerbo, we
must consider together the constraints of fixed energy density $e$, energy
flux $\vec{J}_E$ and particle flow $\vec{J}_N$. If $\vec{J}_E$ and $\vec{J}%
_N $ are taken as independent variables, it is possible to demand that the
particle flow be null in order to eliminate from the energy flux any
advective contribution so that it reduces to a pure heat flux. Notice that
the distribution function that maximizes the entropy can no longer be an
equilibrium one, as there is no reference frame in which one could find an
equilibrium system simultaneously at rest (i.e., with no photon flow) and
with an energy flux. In addition, from the general expressions obtained in
this section, one can recover in the appropriate limit, expressions
previously obtained in the literature for radiation submitted to an energy
flux and to a particle flux.

Let us note that several reasons can induce a non Planckian distribution for
a photon gas coupled to matter in a nonequilibrium state submitted to high
gradients or rapidly varying fluxes. On the one hand, photons can interact
weakly with matter in the time scale over which the flow variables change.
For this reason, the temperature of the photon gas can be different from the
local equilibrium temperature of the matter. In these situations, a
Stefan-Boltzmann-like law, namely $aT_R^4$, may be used with a temperature $%
T_R$ different from the local equilibrium temperature $T_M$ of the matter.
This situation appears for example in the so-called diluted radiation \cite
{Lands2}, of interest in photo voltaic devices. On the other hand, if we
consider matter submitted to large temperature gradients and the length
scale of its interaction with photons is large in comparison to the scale of
variation of temperature in matter, the distribution of photons can be
anisotropic. The previous examples are attempts to include these effects in
the statistical distribution of photons and in the Eddington factor.

The distribution function to consider will then be 
\begin{equation}
n(\vec{p})=\frac 2{\left[ \exp \left( \beta pc+\vec{I}\cdot pc\vec{c}+\vec{K}%
\cdot \vec{c}\right) -1\right] }\approx 2\ \exp \left[ -\beta pc-\vec{I}%
\cdot pc\vec{c}-\vec{K}\cdot \vec{c}\right]  \label{ngeneral}
\end{equation}
where $\vec{I}$, $\vec{K}$ are the Lagrange multipliers related to energy
flux and particle flux respectively and the factor $2$ is related to the two
possible polarizations of photons. If $\vec{I}=0$ the constrained flux is
the particle flux and one recovers the results obtained by Minerbo \cite
{Minerbo}, while when one takes $\vec{K}=0,$ the constrained flux is the
energy flux and the Lorentz limit is recovered. Any integral quantity
defined in terms of this latter approximated distribution converges provided
that $\beta >|\vec{I}|c$. In addition, in order to simplify the
calculations, we will assume that $\vec{I}$ and $\vec{K}$ are parallel. With
these requirements we can find any thermodynamic quantity in terms of the
special functions defined by: 
\begin{equation}
\Psi _n(a,b)\equiv \int_{-1}^1\frac{e^{-ax}}{(1+bx)^n}dx,
\end{equation}
which verify the useful properties: 
\begin{equation}
\frac{\partial \Psi _n}{\partial a}=\frac 1b\left[ \Psi _n-\Psi
_{n-1}\right] ,
\end{equation}
\begin{equation}
\frac{\partial \Psi _n}{\partial b}=\frac nb\left[ \Psi _{n+1}-\Psi
_n\right] .
\end{equation}
In particular, the Lagrange multipliers $\vec{K}$, $\vec{I}$ can be related
with the dissipative fluxes by using (\ref{ngeneral}) in Eqs.(\ref{Je}),(\ref
{fluxN}), leading to new nonequilibrium equations of state: 
\begin{equation}
J_E=\frac{24\pi c}{(hc)^3\beta ^4}\frac 1b\left[ \Psi _3(a,b)-\Psi
_4(a,b)\right] ,  \label{Jegeneral}
\end{equation}
\begin{equation}
J_N=\frac{8\pi c}{(\beta hc)^3}\frac 1b\left[ \Psi _2(a,b)-\Psi
_3(a,b)\right] ,  \label{Jvelocitat}
\end{equation}
where we have defined $a:=|\vec{K}|c$ and $b:=|\vec{I}|c/\beta $, while the
caloric equation of state can be computed from (\ref{e}) and is given by:
\begin{equation}
e=\frac{24\pi }{(hc)^3\beta ^4}\Psi _4(a,b).  \label{egeneral}
\end{equation}
Note that the presence of dissipative fluxes is seen to modify the
Stefan-Boltzmann law. The density of photons can be also computed from (\ref
{densitat}) and is given by:
\begin{equation}
\rho =\frac{8\pi }{(hc\beta )^3}\Psi _3(a,b).  \label{rhogeneral}
\end{equation}
The reduced fluxes $f_E$ and $f_N$ are given by:
\begin{equation}  \label{fegeneral}
f_E=\frac 1b\left[ \frac{\Psi _3(a,b)}{\Psi _4(a,b)}-1\right] ,
\end{equation}
\begin{equation}
f_N=\frac 1b\left[ \frac{\Psi _2(a,b)}{\Psi _3(a,b)}-1\right] ,
\label{fngeneral}
\end{equation}
so they do not coincide in general. The Eddington factor calculated from (%
\ref{Pe}) and (\ref{Edfact}) is given by:
\begin{equation}
\chi =\frac 1{b^2}\left[ 1-2\frac{\Psi _3(a,b)}{\Psi _4(a,b)}+\frac{\Psi
_2(a,b)}{\Psi _4(a,b)}\right] =\frac 1b\left[ f_N-f_E+bf_Nf_E\right] .
\label{Edgeneral}
\end{equation}
and, thus, it is given in parametric form as a function of both fluxes, $f_E$
and $f_N$ by Eqs. (\ref{fegeneral}),(\ref{fngeneral}) and (\ref{Edgeneral}).
The behavior of this Eddington factor for low flux values is analyzed below,
but let us note that in equilibrium ($f_N=f_E=0$), the isotropic Eddington
factor $\chi =1/3$ is recovered. Now, we discuss Eq. (\ref{Edgeneral}) in
some important particular cases.

We start by considering the situation of a pure heat flow, which corresponds
to energy transport without net mass flow, i.e. we impose that $f_N=0$. From
Eq. (\ref{Jvelocitat}), we can easily observe that this condition is simply
given by: 
\begin{equation}
\Psi _2(a,b)=\Psi _3(a,b),  \label{v0}
\end{equation}
thus $a$ and $b$ are no longer independent variables. Using equations (\ref
{fegeneral}) and (\ref{v0}) one can express $a$ and $b$ as a function of $%
f_E $. In this case of pure heat flux, the Eddington factor adopts a simpler
form,
\begin{equation}
\chi =\frac 1{b^2}\left[ 1-\frac{\Psi _2(a,b)}{\Psi _4(a,b)}\right] =-\frac{%
f_E}b.  \label{Edfluxnul}
\end{equation}
Here we recall that first the photon flux $f_N$ is constrained and then is
settled to $0$ {\it a posteriori}.

If the photon flux $f_N$ is unconstrained, which corresponds to the limit $%
a=0$, $f_N$ is related to $f_E$ by:
\begin{equation}
f_N=\frac 1{f_E}(2-\sqrt{4-3f_E^2}),
\end{equation}
and one recovers the so-called Lorentz Eddington factor, namely 
\begin{equation}
\chi =\frac 53-\frac 23\sqrt{4-3f_E^2.}  \label{Lorentz}
\end{equation}

This Eddington factor was proven to be corresponding to an equilibrium
moving system in \cite{MiniCuca}: the constraint of given energy flux $\vec{J%
}_E$ (or equivalently, given momentum $\vec{P}$, as $\vec{J}_E=c^2\vec{P}$)
can be realized with a (local-)equilibrium moving system if one does not
constrain the particle flow of the system.

In the limit $b\rightarrow 0$ (fixed particle flux $\vec{J}_N$ but
unconstrained $\vec{J}_E$) we obtain a simpler expression for the functions $%
\Psi _n\left( a,b=0\right) $ 
\begin{equation}
\Psi _n\left( a,b=0\right) =\int_{-1}^1\exp \left( -ax\right) dx=2\frac{%
\sinh a}a,
\end{equation}
which used in Eqs. (\ref{Jegeneral}), (\ref{Jvelocitat}), (\ref{egeneral}), (%
\ref{rhogeneral}) and (\ref{Edgeneral}) leads to the parametric Eddington
factor:
\begin{eqnarray}
f_E &=&f_N=\frac 1a-\coth a,  \label{Minerbo} \\
\chi &=&1-\frac 2a\left( \frac 1a-\coth a\right) .  \nonumber
\end{eqnarray}
Eqs. (\ref{Minerbo}) were obtained previously by Minerbo \cite{Minerbo} in
one of the first attempts to obtain a variable Eddington factor for
anisotropic nonequilibrium radiation from probabilistic arguments.
Contrarily to the Lorentz case, it corresponds to a true nonequilibrium
situation: the distribution of photons (\ref{ngeneral}) with $b=0$ cannot be
transformed into a Planckian by a Lorentz transformation.

In Figure \ref{figura} we compare the Eddington factors (\ref{Lorentz}), (\ref
{Minerbo}), (\ref{Edfluxnul}) and an expression arising from a
Chapmann-Enskog calculation and expressed in a parametric form by \cite
{Lever}: 
\begin{eqnarray}
f_E &=&1/m-\coth m, \\
\chi &=&-\coth m\cdot (1/m-\coth m)  \nonumber
\end{eqnarray}

Note that the pure heat flux case, Eq. (\ref{Edfluxnul}) grows with the
energy flux more rapidly than those of Lorentz and Minerbo as any advective
contribution to the energy flux has been subtracted by demanding that $%
f_N=0. $

The entropy density is also modified due to the external fluxes, yielding 
\begin{equation}
s(e,J_E,J_N)=\frac{8\pi }{(hc\beta )^3}k_B\left[ \left( 4-\frac ab\right)
\Psi _3(a,b)+\frac ab\Psi _2(a,b)\right] .  \label{entropy}
\end{equation}
In the limit of vanishing fluxes, $a\rightarrow 0$, $b\rightarrow 0$ we
recover the equilibrium entropy multiplied by a factor $90/\pi ^4\simeq 0.92$
which is due to the approximation of neglecting the $-1$ term in the
distribution function (\ref{ngeneral}).

In the case of pure heat flux Eq. (\ref{entropy}) takes a more simpler form.
Using equations (\ref{Jegeneral}) and (\ref{v0}) one can express $a$ and $b$
as a function of $J_E$ and we can write for the entropy
\begin{equation}
s(e,J_E)=\frac{32\pi }{(hc\beta )^3}k_B\Psi _3(a,b)=4\,k_B\,\rho .
\end{equation}
The relation coming from the last equality is also known to hold in
equilibrium.

The nonequilibrium entropy density (\ref{entropy}) in the Minerbo's limit $%
b=0,$ is given by 
\begin{equation}
s\left( e,\vec{J}_N\right) =\frac{45}{2\pi ^4}s_{eq}\left( e\right) \left( 
\frac a{\sinh a}\right) ^{3/4}\left[ 5\frac{\sinh a}a-\cosh a\right] ,
\end{equation}
and the energy and photon densities are given by 
\begin{eqnarray}
e &=&\frac{48\pi }{\left( hc\right) ^3\beta ^4}\frac{\sinh a}a, \\
\rho &=&\frac{16\pi }{\left( hc\beta \right) ^3}\frac{\sinh a}a,  \nonumber
\end{eqnarray}
Note that, the larger the flux, the lower the entropy, revealing the larger
order existing in the system. In the limit $a\rightarrow \infty ,$ i.e. all
photons moving collectively, the entropy $s\rightarrow -\infty $ and $\beta
\rightarrow \infty ,$ while in the limit $a\rightarrow 0$ we recover the
equilibrium situation.

Once the nonequilibrium entropy density is known, the thermodynamic pressure
can be obtained from the relation :
\begin{equation}
s=\frac eT+\frac pT+k_B\vec I\cdot \vec J_E+k_B\vec K\cdot \vec J_N,
\end{equation}
(being $T=1/k_B\beta $) and it verifies:
\begin{equation}
p=\frac{8\pi }{(hc)^3\beta ^4}\Psi _3(a,b)=\frac \rho \beta ,
\label{pgeneral}
\end{equation}
as in equilibrium. However, now the equilibrium relation between the
thermodynamic pressure $p$ and the trace of the pressure tensor, namely $Tr(%
{\bf P}_E)=3p,$ is not satisfied. This can be shown by noting that the trace
of the pressure tensor is $Tr({\bf P}_E)=e$ for relativistic particles, and
that the relation between the thermodynamic pressure $p$ and the energy
density $e$ as obtained from Eqs. (\ref{egeneral}),(\ref{pgeneral}),
\begin{equation}
p=\frac e3\frac{\Psi _3(a,b)}{\Psi _4(a,b)}=\frac e3+\frac b{3c}J_E,
\label{presio}
\end{equation}
differs from the equilibrium one, $p=e/3$. From Eq. (\ref{presio}), one
obtains that the relation $p=e/3$ holds both in the Minerbo's case ($b=0$)
and in equilibrium at rest ($J_E=0$).

The thermodynamic pressure $p$ is not related either to the isotropic part
of the pressure tensor, namely $e(1-\chi )/2$, in the general situation as
shown by inspection of Eqs.(\ref{egeneral}), (\ref{Edgeneral}) and (\ref
{pgeneral}): $p=e(1-\chi )/2$ only in the Lorentz's case and in equilibrium
at rest.

If we restrict ourselves up to second order in $\vec{J}_E$ and $\vec{J}_N$,
we can find simpler analytical expressions for the thermodynamic quantities
than the previous ones, what allows a simpler physical interpretation.
First, a simple relation between the Lagrange multipliers and the fluxes is
seen to hold:
\begin{equation}
f_N=\;-\frac a3-b,\qquad f_E=-\frac a3-\frac{4b}3,
\end{equation}
and we can, thus, obtain an explicit flux-dependent caloric equation of
state up to second order:
\begin{equation}
\beta \simeq \beta _0\left[ 1+\frac 38\left( 5f_E^2-8\vec{f}_N\cdot \vec{f}%
_E+4f_N^2\right) \right] ,\quad \mbox{with }\quad \beta _0:=\left[ \frac{%
24\pi }{(hc)^3e}\right] ^{1/4}.  \label{betaaprox}
\end{equation}
Note that $T_0=1/k_B\beta _0$ is the temperature corresponding to
equilibrium radiation with the same energy density $e$. Thus, Eq. (\ref
{betaaprox}) shows that the nonequilibrium mean temperature defined as $%
T=1/k_B\beta $ is smaller than the corresponding local equilibrium
temperature $T_0$, as expected from the general arguments of Landau \cite
{Landau} which must hold for all possible nonequilibrium distributions of
radiation.

The entropy density of the system can also be written as the equilibrium one
plus a quadratic correction in the fluxes:
\begin{equation}
s(e,J_E,J_N)=\frac{64\pi }{(hc\beta _0)^3}k_B\left[ 1-\frac 32\left( \frac 34%
f_E^2-\frac 32\vec f_N\cdot \vec f_E+f_N^2\right) \right] ,
\end{equation}
and, as expected, the flux dependent contribution reduces the value of the
entropy, in agreement to the fact that it corresponds to a more ordered
physical situation.

The particle density verifies
\begin{equation}
\rho =\frac{16\pi }{(hc\beta _0)^3}\left[ 1-\frac 32\left( \frac 34%
f_E^2-f_N^2\right) \right] ,
\end{equation}
and the Eddington factor can be approximated by 
\begin{equation}
\chi \simeq \frac 13+\frac 25\left[ 5f_E^2-8\vec{f}_E\cdot \vec{f}%
_N+4f_N^2\right] .  \label{Edap}
\end{equation}
Let us observe that both fluxes must be null in order to recover the
isotropic Eddington factor $\chi =1/3$, due to the fact that the fluxes $%
f_N, $ $f_E$ are considered as independent variables in the derivation. This
situation is different from those encountered in the Eddington factor
depending on two parameters introduced in \cite{Fu}, where $f_N$ and $f_E$
are considered as dependent variables and the Eddington factor depends on $%
f_E$ and a nonequilibrium chemical potential for the photons (which must be
set by hand to zero at equilibrium).

We should also notice that the correction to the isotropic Eddington factor
is always positive (the bilinear form is always positive) up to second order
in the fluxes. In addition, from Eq. (\ref{Edap}) we can observe that the
case of pure heat flux ($f_N=0$) corresponds, among the situations we have
considered, to the case in which the correction to the isotropic value $1/3$
due to $f_E$ is higher: from Eq. (\ref{Edap}) it is seen to be $4$ times
higher than that observed in the Lorentz situation ( $f_N=3/4f_E$) and $5$
times higher than that of Minerbo ($f_N=f_E$).

\section{Anisotropic radiation with chemical potential}

Up to now, we have considered, as usual, that photons have a null chemical
potential, so its number is undetermined. However, the use of a non-zero
chemical potential for a photon gas has already been proposed in several
(and physically distinct) situations. For example, in solid state physics,
non-equilibrium but steady-state quasi-Fermi distributions had been used
since the 1950's for the electron gas. In this case, the study of a photon
gas coupled with this electron gas via adsorption-emission processes leads 
\cite{Lands2} to a Bose-Einstein distribution 
\begin{equation}
n(\epsilon )=\frac 2{\exp ({\beta \epsilon +\mu _{ph}\beta )}-1},
\end{equation}
where $\mu _{ph}$ is the chemical potential of the photons. Such a non-null
chemical potential is related to stimulated emission of photons and other
out of equilibrium situations. In astrophysics, massless particles like
neutrinos and photons have been considered with nonzero chemical potential
in nonequilibrium situations, when the number conserving Compton scattering
dominates (see for example \cite{Fu} and references therein).

From a technical point of view, the occupation number given in (\ref
{ngeneral}) leads to divergent integrated quantities out of the
low-occupation number approximation. This drawback can also be removed by
introducing a nonequilibrium chemical potential. In addition, if we consider
a photon gas with energy density $e$ which transports an energy flux $\vec{J}%
_E$ with a fixed particle flow $\vec{J}_N$, it seems quite reasonable to
impose together with these constraints to the entropy, the constant mean
number of particles $\rho $, being this latter imposed in order to be
consistent with the fact of imposing a fixed value for the particle flow.
This assumption implies, as in the examples considered above, that photons
out of equilibrium may have a non null chemical potential. The occupation
number which maximizes the entropy (\ref{Shannon2}) under these constraints
is: 
\begin{equation}
n=\frac 2{\exp ({\beta pc+\mu _{ph}\beta +\vec{I}\cdot pc\vec{c}+\vec{K}%
\cdot \vec{c})}-1}.
\end{equation}
Once the distribution function is known, one can obtain the thermodynamics
of the system. As above, we restrict ourselves to the low occupation number
limit in order to compare with the previous results. This limit corresponds
to an ultrarelativistic classical gas (apart from the factor $2$ related to
the polarization of photons). Following the same procedure as in the
previous sections, we can easily obtain in this approximation (assuming as
in the previous section, that $\beta >|\vec{I}|c$ so that integrations
converge and that $\vec{I}$ and $\vec{K}$ are parallel for simplicity). With
these requirements we can find ($a\equiv |\vec{K}|c$ and $b\equiv |\vec{I}%
|c/\beta $): 
\begin{equation}
\rho =\frac{8\pi }{(hc\beta )^3}\exp (-\mu _{ph}\beta )\Psi _3(a,b),
\end{equation}
\begin{equation}
e=\frac{24\pi }{(hc\beta )^3\beta }\exp (-\mu _{ph}\beta )\Psi _4(a,b),
\label{enu}
\end{equation}
\begin{equation}
J_E=\frac{24\pi c}{(hc\beta )^3\beta }\exp (-\mu _{ph}\beta )\frac 1b\left[
\Psi _3(a,b)-\Psi _4(a,b)\right] ,
\end{equation}
\begin{equation}
J_N=\frac{8\pi c}{(hc\beta )^3}\exp (-\mu _{ph}\beta )\frac 1b\left[ \Psi
_2(a,b)-\Psi _3(a,b)\right] .  \label{velocitat}
\end{equation}

On the other hand, the entropy density can be written as 
\begin{equation}
s=k_B\rho \left[ \ln \left( \frac{8\pi }{(h\beta c)^3\rho }\right) +4+\ln
\Psi _3(a,b)+\frac ab\left( \frac{\Psi _2(a,b)}{\Psi _3(a,b)}-1\right)
\right] ,
\end{equation}
and the thermodynamic pressure is also given by Eq. (\ref{presio}).

With the exception of pressure, all these quantities differ from the ones
introduced in the previous section. However the reduced variables, namely $%
f_E,$ $f_N$ and $\chi $ are the same, as the new contribution due to
chemical potential vanishes. Therefore, this new approach does not affect
the study of radiative transfer, though it clearly modifies the
thermodynamics of the system. Note, however, that the new non null
nonequilibrium chemical potential for photons can be written, in terms of
the particle density of the system as
\begin{equation}  \label{mu}
\mu _{ph}=\frac 1\beta \ln \left( \frac{8\pi }{\rho (h\beta c)^3}\Psi
_3\right)
\end{equation}
so it does not reduce to zero for null fluxes, unless we also require
independently that $\rho $ be given by the equilibrium expression $\rho
=16\pi /(h\beta c)^3$. This fact arises directly from the hypothesis that a
fixed particle density can be fixed independently from the fluxes. However,
this is also the case, for instance, in the treatment performed by Fu \cite
{Fu}.

In order to compare with the results in the previous section, we can rewrite
the particle density $\rho $ as 
\begin{equation}
\rho =\frac{8\pi }{(h\beta c)^3}\Psi _3(a,b)\left( 1+\alpha \right) ,
\end{equation}
where $\alpha $ is the modification with respect to the particle density in (%
\ref{rhogeneral}), so the rest of the previous quantities can also be found
to be given by 
\begin{equation}
e=\frac{24\pi }{(h\beta c)^3\beta }(1+\alpha )\Psi _4(a,b),
\end{equation}
\begin{equation}
J_E=\frac{24\pi c}{(h\beta c)^3\beta }(1+\alpha )\frac 1b\left( \Psi
_3(a,b)-\Psi _4(a,b)\right) ,
\end{equation}
\begin{equation}
J_N=\frac{8\pi c}{(h\beta c)^3}(1+\alpha )\frac 1b\left( \Psi _2(a,b)-\Psi
_3(a,b)\right) ,
\end{equation}
so that, if the chemical potential is not null, the thermodynamic equations
of state introduced in Section 3 will be modified by a factor $\left(
1+\alpha \right) $.

\section{Conclusions}

In this paper we have applied information theory based nonequilibrium
statistical mechanics to describe anisotropic radiation in the so-called
variable Eddington factor scheme, investigating the main thermodynamic
features of the system.

Typically, in the variable Eddington factor closure scheme, one considers
the reduced energy flux $\vec f_E$ as a new variable and the pressure tensor
is written as a function of it to close the hierarchy of radiative transfer
equations. However, this procedure, widely used in the applications of the
radiative transfer equation, has some thermodynamic consequences which are
usually skipped. The main consequence is that the local equilibrium
hypothesis must be abandoned as it is not expected to provide any dependence
in the energy flux for the pressure tensor except in the case in which we
consider radiation in a moving frame. The local equilibrium hypothesis
implies a local black body distribution for radiation, but it is known \cite
{Essex} that in many radiation transfer problems the photon distribution
strongly departs from this simpler black body expression.

The use of information theory allows one to obtain a distribution function
for the photon gas consistent with this closure scheme, and the
flux-dependent Eddington factor as well as the thermodynamics of the system
can be obtained. In this formalism, the flux dependence of the Eddington
factor merely reflects the fact that the fluxes characterizing the
anisotropy of the system are incorporated as a new thermodynamical
variables. The necessity of incorporating dissipative fluxes within the set
of thermodynamic variables beyond the local equilibrium hypothesis has been
introduced by recent nonequilibrium thermodynamic theories \cite
{Ext,Ext2,Ext3}.

In the variable Eddington closure scheme we have two fluxes in the system:
the energy flux $\vec{J}_E$ and the particle flux $\vec{J}_N$. If these
moments are taken as dependent variables, we must use only one of them as a
constraint in the informational entropy. The effect of considering each of
them as a constraint has already been performed \cite
{Minerbo,Anile,Kremer,Inf}, whereas the effect of considering them as
independent variables is the subject of Sections 2 and 3.

If the energy flux $\vec J_E$ is taken as a constraint in the maximization
of the informational entropy, the results obtained by a purely macroscopic
method in \cite{Anile,Kremer} are recovered. As already pointed in \cite
{MiniCuca}, the results obtained in \cite{Anile,Kremer,Inf} were not really
a description of nonequilibrium radiation, but merely equilibrium radiation
as observed from a moving reference frame. Thus, these procedures allow the
study of anisotropic radiation but, being the anisotropy due to the relative
motion between matter and radiation (like in the case of the cosmic
background radiation), they do not lead to a nonequilibrium situation. Both
information theory and the macroscopic approach followed in \cite
{Anile,Kremer} consider an entropy depending on the energy flux but no
restrictions about the global motion of the system are imposed. Therefore,
when the condition of maximum entropy is used, an equilibrium moving system
appears because equilibrium situations have the maximum entropy and the
moving system verifies the imposed constraint of non zero energy flux. In
fact, irreversibility is related to positive entropy production and this
fact, intrinsically related to the nature of the processes occurring in the
system, cannot be changed by simply performing a Lorentz boost.

Nevertheless, from the point of view of nonequilibrium thermodynamics beyond
local-equilibrium, we can observe that, according to the results in \cite
{Inf}, the existence of a flux-dependent temperature might seem a physically
reasonable assumption in this context, as it arises from a purely
equilibrium situation. This would suggest a further validity of
flux-dependent equations of state for a more general, nonequilibrium
situation.

To obtain a true nonequilibrium situation, one can impose the particle flux $%
\vec{J}_N$ as a constraint, instead of the energy flux $\vec{J}_E$ as done
by Minerbo \cite{Minerbo}. His definitions were frequency-dependent and ours
are not. However, if we consider a gray medium and the radiation intensity
factorizes into a frequency-dependent and an angular dependent part,
Minerbo's Eddington factor can be recovered applying information theory
(with integrated frequencies) submitted to a fixed particle flow.

In Section 2, by introducing both fluxes $\vec{J}_E$, $\vec{J}_N$ as
independent variables, we have obtained a unified treatment that allows the
re-derivation of both Lorentz's and Minerbo's cases in the proper limits. In
addition, we obtain new more general forms for the (two-parameter) Eddington
factor that may be useful when the matter velocity is not null with respect
to the photon flow and such an effect introduces an advective contribution
to the energy flux. Hence, we have considered the Eddington factor
describing radiation under a pure heat flow (particle flow is set to zero),
for which the anisotropic effects due to the flux seem to be higher than in
any of the previous situations.

In addition to nonequilibrium Eddington factors, we have also studied the
nonequilibrium equations of state arising from such formalism. Note that,
whereas the equation 
\begin{equation}
p=\rho k_BT
\end{equation}
always remains, the nonequilibrium temperature $T$ is related to energy by:
\begin{equation}
\frac 1T=k_B\left( \frac{24\pi }{(hc)^3e}\Psi _4(a,b)\right) ^{1/4},
\label{noneqt}
\end{equation}
so a nonequilibrium, flux-dependent temperature appears. This fact is in
complete agreement with the predictions of some macroscopic theories \cite
{Temp,Temp2,Temp3} that have introduced nonequilibrium equations of state.
Note as well that $\Psi _4(a,b)\geq 1,$ so the nonequilibrium temperature $T$
is smaller than the equilibrium one. With regard to the thermodynamic
pressure $p$, given by 
\begin{equation}
p=\frac e3+\frac{J_Eb}{3c},
\end{equation}
it is not related in the general case neither to the trace of the pressure
tensor, as usual, nor to its isotropic part. This situation is, in fact,
analogous to that encountered in \cite{Inf} for a classical ideal gas
submitted to a heat flux. However, if $b=0$, $p=e/3$ and it is thus related
to the trace of the pressure tensor; whereas in the case of an equilibrium
moving system ($a=0$), $p=e\frac{1-\chi }2,$ i.e. it is given by the
isotropic part of the pressure tensor.

Finally, we have also considered the case in which a nonequilibrium chemical
potential for photons is introduced as suggested in \cite{Lands2} and done
in \cite{Fu}. Although the necessity of such an assumption is not clear, it
seems a plausible ansatz which also removes the technical difficulties of
convergence beyond the low occupation number limit. Within this limit, if $%
\mu _{ph}$ is introduced, the radiative properties of the system are not
modified, but in the thermodynamic equations of state a new factor $%
(1+\alpha )$ appears. The general case deserves further study and will be
the object of a future paper.

\section*{Acknowledgments}

Stimulating discussions with Jordi Faraudo, Professor D. Jou and Professor
J. Casas-V\'{a}zquez from the Autonomous University of Barcelona have played
an important role in this work. The author is supported by a doctoral
scholarship from the Programa de formaci\'{o} d'investigadors of the
Generalitat de Catalunya under grant No. FI/94-2.009. Partial financial
support from the Direcci\'{o}n General de Investigaci\'{o}n of the Spanish
Ministry of Education and Science (grant PB94-0718) is also acknowledged.

\newpage
\begin{figure}[tbp]
\caption{Functional relationship between $\chi$ and $f_E$ for four different
models. Solid line: pure heat flux; dashed line: Levermore (eq. (72)-(73));
short dashed line: Minerbo ($b=0$); dots: Lorentz ($a=0$).}
\label{figura}
\end{figure}

\end{document}